\documentclass{llncs}
\usepackage{llncsdoc}

\title{Grammatic -- a tool for grammar definition reuse and modularity}
\author{Andrey Breslav}
\institute{St. Petersburg State University of Information Technology, Mechanics and Optics
\email{abreslav@gmail.com}
}

\begin{document}
\maketitle
\begin{abstract}
Grammatic is a tool for grammar definition and manipulation aimed to improve modularity and reuse of grammars and related development artifacts.
It is independent from parsing technology and any other details of target system implementation. 
Grammatic provides a way for annotating grammars with arbitrary metadata (associativity attributes, semantic actions or anything else). 
It might be used as a front-end for external tools like parser generators to make their input grammars modular and reusable.
This paper describes main principles behind Grammatic and gives an overview of languages it provides and their ability to separate concerns and define reusable modules. Also it presents sketches of possible use cases for the tool.
\end{abstract}

\section{Introduction}
To follow ideas of language-oriented programming \cite{LanguageOriented} or DSL-based development we need many DSLs, which may mean dozens for a medium-sized project. To be able to produce and support so many languages we need tools which strongly support reuse and modularity. We need to reuse development artifacts to produce many languages since even different ones may sometimes have much in common and it is really hard to work on a bunch of languages having common features implemented by code duplication. The primary condition for reuse is modularity therefore we will always mention them together.

Language development involves many aspects but here we will mostly concentrate on textual syntax. Structure of language syntax is usually described by a context-free grammar. Engineering aspects of grammar development did not yet receive enough attention \cite{KLV05}. To provide tools suitable for industry we need to consolidate and improve results achieved by different authors. We give a brief overview of existing tools in section \ref{RelatedWork}. 

In this paper we describe Grammatic -- a tool which is aimed to provide reuse and modularity for grammar definitions which might be used for arbitrary purpose (not only parser generation but also analysis of grammars, language synthesis, layout and many other applications). Grammatic provides general languages for context-free grammar definition and annotation. We present them in section \ref{Languages}. These languages are designed to support modularity and reuse, we describe these features in section \ref{Reuse}.

 Grammatic is mostly dedicated to be a front-end for other tools and a framework for high-level manipulations with grammars and metadata. As a front-end Grammatic does not replace existing tools but helps to extend and join their powerful abilities. We outline some possible use cases in section \ref{UseCases}.

\section{Related work}\label{RelatedWork}
The best developed and most widely used family of tools working with grammars are parser and compiler generators. Here we give an overview of most popular and powerful of them. Our goal is to present features supporting modularity and reuse, so we concentrate on these aspects here. Ideas implemented in some of these tools significantly influenced our approach.

\subsection{Separation of concerns}
The most significant modularity issues are related to separation of syntactic and semantic information in grammar definitions.

Traditional parser generators (Yacc \cite{yacc}, ANTLR \cite{Parr2007}, COCO/R \cite{Mossenbock05thecompiler} and many others) use grammar definitions with embedded semantic actions. This leads to mixing up syntactical and semantical aspects of the system into one definition file: it lacks modularity and is hard to maintain.

Tools specially dedicated to creating textual syntax for simple DSLs (xText \cite{oAWxText}, TCS \cite{JouaultBezKur2006} and others) use grammar annotations needed for their specific purposes (namely building target models) and thus also mix syntax with semantics. 

SableCC \cite{Gagnon98anobject-oriented} avoids mixing things up: it does not allow to embed semantic actions. A parser generated by SableCC creates an abstract syntax tree (AST) and the user has to write his/her own custom code to analyze it or transform into the target format. This provides a clear separation of concerns.

SDF \cite{BDHJJKKMOSVVV01} also separates semantic actions from parsing by restricting parser responsibilities to building a tree. Unlike SableCC, SDF provides much more powerful grammar definition language.

Another approach to separation of concerns is presented by LISA compiler-compiler \cite{Mernik00compiler}: it provides an opportunity to either embed semantic actions or attach them using AOP-like point cuts. This allows to establish different modules (aspects) even for semantic actions themselves which improves overall system modularity.

\subsection{Reuse}
A basic way to reuse grammar definitions in textual form is to use a C-like preprocessor (any system taking textual input allows reuse with preprocessor). This is the only way to go when using many traditional tools.

One of the most popular parser generators today is ANTLR. In version 2 it provided ``grammar inheritance'' with ``rule overriding'' and in latest version 3.1.1 it introduced grammar imports \cite{Parr08} which are rather restricted.

SDF has a powerful import mechanism, supports grammar composition and templates (parameterized modules).

LISA allows grammar inheritance, aspect-oriented composition and templates for semantic actions.

\section{Grammatic's languages overview}\label{Languages}
Grammatic defines a language for describing context-free grammars which does not involve anything but pure grammar productions. This language supports EBNF constructs for both lexical and syntactical rules.

Grammar is not enough to build the whole system. We need some semantics-related or other information to be added to the grammar to have a complete system description. For this purpose Grammatic defines a languages for attaching metadata to grammars and their elements.

Metadata annotations are grouped into aspects. A metadata aspect can be though of as a description of the system made from a certain point of view: parsing, pretty printing, building target objects etc.

Below we describe these languages in more details.

\subsection{Grammar definitions}
A grammar definition language allows to define a grammar as a set of symbols each of which is associated to a set of productions (in concrete syntax we separate productions by ``$|$$|$''). Grammatic itself does not restrict grammars to any particular class (like LL(k), LALR or anything else). Here is an example from a grammar for arithmetic expressions:

\begin{verbatim}
Factor
    -> Literal
    || ID
    || '(' Expression ')'
    ;
\end{verbatim}

In this example characters in single quotes represent embedded lexical definitions. There is no separate notion of a lexical rule (since it is not necessarily required, see \cite{BDHJJKKMOSVVV01}) and all the regular expression operations (sequence, alternative, iteration) are available both for syntactical and lexical definitions. Here is a syntactic (nonterminal) symbol:
\begin{verbatim}
Product 
    -> Factor ('*' Factor)*
    ;
\end{verbatim}

And these are ``lexical'' (terminal) symbols:
\begin{verbatim}
INT
    -> ['0'--'9']+
    ;
REAL
    -> INT ('.' INT)? (('e' | 'E') ('+' | '-')? INT)?
    ;
\end{verbatim}

Separation of lexer and parser is not needed in some cases (see \cite{BDHJJKKMOSVVV01}). By default lexical symbols do not differ from syntactical. If it is needed, we can can specify a symbol as lexical by providing corresponding metadata.

\subsection{Metadata}
Pure grammar definition defines a textual form of a language but there are many more things to express. Existing tools (see section \ref{RelatedWork}) use some extensions to grammar definitions: embedded semantic actions, AST building instructions etc. Such things might be generally described by metadata attached to a grammar. Grammatic allows to attach arbitrary metadata to a grammar definition.

Metadata annotations might be attached to a grammar, symbol, individual production or a subexpression.
Each annotation may contain several attributes (name-value pairs). 
Attribute values may be of different types. There are several predefined value types:
identifier, string, integer, annotation and sequence of values and punctuation symbols. 

\begin{verbatim}
  id = someName;       // the value is an identifier
  str = "some string"; //                 string
  int = 10;            //                 integer
  class = {            //                 annotation 
      name = MyClass; 
      super = Object;
  };          
  astProduction = {{   //                 sequence
      ^('+' left ^('-' right 10)) 
  }}; 
\end{verbatim}

Users may add their own types. No attribute itself has any fixed semantics. Metadata is passive, some tools (like analyzers, transformers, generators etc.) may use it according to their needs.

Even without adding custom types many things might be expressed by such annotations. The most powerful type is a sequence of values -- it allows to define small embedded DSLs inside Grammatic. We use such DSLs to describe complicated custom properties when working with external tools (i.e. high-level definitions for AST structures).

\subsection{Queries}

How to attach metadata to a grammar? In many cases it is done by directly embedding annotations into grammar definition. Therefore different concerns are mixed together and this results into a problem: system is not modular, is hard to understand and extend.

We employ ideas from aspect-oriented programming paradigm (AOP, see \cite{Kiczales97aspect-orientedprogramming}) to solve this problem. We consider experience of creators of LISA \cite{Mernik00compiler} and extend their ideas. In Grammatic a grammar definition itself knows nothing about metadata. All the metadata is attached ``from the outside''. In AOP this is done by defining join points which are described by point cuts \cite{680006}. A language of point cuts is a kind of ``addressing'' notation -- a way to find some object. In original AOP this object may be a class or a method, in our case this will be a rule, subexpression or another object from a grammar definition. When we have found such an object we may attach metadata to it (or perform other actions, see below).

In Grammatic we have a language analogous to AOP point cuts -- we call it a query language. We use structural queries with embedded variables. For example this query matches rules defining binary operations:

\begin{verbatim}
Op -> Arg (Sign Arg)* ;
\end{verbatim}

All the names here represent variables. This query matches any of the following rules:

\begin{verbatim}
Product
    -> Factor (MultOrDiv Factor)*
    ;
Sum
    -> Product (PlusOrMinus Product)*
    ;
\end{verbatim}

By default a variable matches a symbol but it may match a subexpression or a whole production.

\begin{verbatim}
A -> Alt=(B | C);
\end{verbatim}

Here variables A, B and C match symbol references and Alt matches a subexpression ``B $|$ C''.

We can use wildcards in queries. The following query matches immediately left-recursive rules:

\begin{verbatim}
Rec -> Rec ..;
\end{verbatim}

Two dots represent a wildcard which matches arbitrary subexpression. It also might be assigned to a variable:

\begin{verbatim}
Rec -> Rec Rest=..;
\end{verbatim}

We can consider metadata in our queries. We can restrict a particular attribute to a certain type or value or require attribute's presence or absence:
\begin{verbatim}
N {
    type = Nonterminal;
    operation;
    associativity : ID;
    !commutative;
}
\end{verbatim}

This query matches a symbol with ``type'' attribute having value ``Nonterminal'', ``operation'' attribute present, ``associativity'' attribute having value of type ID and ``commutative'' attribute not present.

\subsection{Aspects}

When a query selects some objects from a grammar definition, we can attach some metadata to them.

\begin{verbatim}
Rec -> Rec ..;
{
    Rec {
        leftRecursive;
    };
}
\end{verbatim}

This rule adds a ``leftRecursive'' attribute (with no value) to all the symbols matched by Rec variable of this query. A set of such rules constitutes an aspect. Many aspects (independent or not) might be assigned to a single grammar, and even to many grammars since our queries are not tied to concrete objects but only to a grammar structure.

\subsection{More applications of queries} \label{MoreQueries}

Query language appears to be useful not only in attaching metadata to grammars. As in AOP (AspectJ, for example) we can use it to define some constraints for our grammars and their metadata. It's done by defining a query and assigning an error message to it: when a query is matched an error (or warning) will be generated. The same might be defined for a case when a query does not match anything on a whole grammar.

Also we can use queries as a general mechanism for locating objects in a grammar which is useful when defining grammar transformations or some generators taking a grammar as input. Here we can consider our query language as analogous to XPath in the context of XSLT.
\section{Reuse}\label{Reuse}

One of the main purposes of Grammatic is improving reuse experience of grammar-oriented tools. An approach described above gives a good basis for it since no concerns are to be mixed together -- we can describe each of them in a separate aspect.

\subsection{Reusing aspects}
Aspects themselves might be generally reusable -- as we told above, query language does not require ``hard linking'' to grammar objects, these objects are located by their structural context and properties. Above we gave an example of marking all immediately left-recursive rules with a ``leftRecursive'' attribute. This is an example of a reusable aspect -- we can use it on any grammar. Although this technique is not very powerful since we frequently define queries which simply describe whole rules with no generalization, it is still useful.

Below we describe more powerful reuse instruments defined by Grammatic.

\subsection{Grammar imports and templates}
First we focus on reusing grammar definitions themselves. There are many achievements in this field done by creators of LISA, SDF and other tools (see section \ref{RelatedWork}).

The most popular way of reuse is importing. Some grammar definition A might be imported into some other grammar definition B. This means that all the rules of A are inserted into B. Rules of B may refer to symbols of A -- this is the way two grammar definitions are connected. 

Very frequently we have to customize some of the imported rules, i.e. add some more productions to the same symbols or replace existing productions. In paper \cite{Parr08} this is referred to as rule overriding. In Grammatic we decided to use more general form of this concept, namely templates.

A language of grammar templates allows creating grammar definitions with ``placeholders'' which can be replaced with actual objects upon template instantiation. Placeholders might be defined for roles of identifier, expression, production or symbol. A template instantiation might result into grammar object of the type specified by template declaration. Any objects except symbols might be used immediately in rule definitions. Symbols are treated as imported from a template instantiation. This is due to naming reasons: to avoid name duplication a template instantiation expression is a namespace and symbols from that namespace might be referred to by qualified names. Here is an example of a template and its usage. 

\begin{verbatim}
Symbol binaryOperation<ID $name, Expression $sign, Expression $argument> {
    $name -> $argument ($sign $argument)*;
}

import binaryOperation<Product, '*' | '/', Factor>;
import binaryOperation<Sum, '+' | '-', Product>;
Factor
    -> NUMBER
    || ID
    || '(' Sum ')'
    ;
\end{verbatim}

In this example we define a template named ``binaryOperation'' which makes up an infix binary operation out of symbol name, sign and argument expression. Then we instantiate it twice and import instantiation results into current grammar definition -- so we can use new symbol ``Product'' to create ``Sum'' and ``Sum'' to define ``Factor''. Here we did not need to use qualified names explicitly -- there's no name duplication.

Now let's assume that we need to refer to signs of our binary operations as separate symbols. We modify the template as follows:

\begin{verbatim}
Symbol binaryOperation<ID $name, Expression $sign, Expression $argument> {
    Sign -> $sign;
    $name -> $argument (Sign $argument)*;
}
\end{verbatim}

Now we get two symbols out of a single template instantiation: one symbol for operation and another (named ``Sign'') -- for its sign. To refer to these new symbols we need qualified names (and named namespaces):

\begin{verbatim}
import product = binaryOperation<Product, '*' | '/', Factor>;
import sum = binaryOperation<Sum, '+' | '-', Product>;

AnySign
    -> product.Sign | sum.Sign
    ;
\end{verbatim}

How can we use templates for ``overriding'' things? We can put a customizable set of rules into a template, provide a placeholder for production or subexpression that should be replaced and then put a right thing in upon instantiation.

\begin{verbatim}
Symbol attributeValue<Production* $moreValueTypes> {
    AttributeValue
        -> STRING
        || ID
        || INT
        || Annotation
        || ValueSequence
        || $moreValueTypes
        ;
}

import attributeValue<
    '{{{' Expression '}}}'
>;
\end{verbatim}

This defines a template for ``AttributeValue'' symbol and instantiates it adding a new production (to use expressions as attribute values).

\section{Use cases}\label{UseCases}
Grammatic might be used for high level manipulation on modular grammar definitions annotated with arbitrary metadata. In practice this means that we can use it as a front-end for some external tools to provide modularity and reuse.

For example, if we need to use Yacc to generate parsers we can write a module which transforms a Grammatic grammar to a Yacc grammar to use Grammaic's modularity and Yacc's generator together. 

But pure grammar will not allow us to generate any meaningful parser. We need to provide some additional information to generate Yacc's semantic actions (and some other stuff which we will not consider here). The easiest way to do it is to provide string-valued attributes for Grammatic productions which simply contain semantic actions' code.
By doing so we will get modular grammar separated from semantic actions which might be transformed into a working parser using Grammatic-to-Yacc generator and then Yacc.

Another way here is to restrict functionality of the parser we are going to get and go after SDF or xText. To do so we must define attributes which describe needed actions in a declarative way: create node of this type, attach this node to this etc. And again we get a working parser by applying two transformations: Grammatic-to-Yacc and then Yacc. Although the first will be more complicated this time.

Note that Grammatic-to-Yacc transformation is to be written once and shipped as a pluggable module for Grammatic. And this might be done for many other external tools (not only parser generators). 

In many cases we need to restrict an input of Grammatic-to-Something transformation: we have to ensure attribute value types, prohibit some constructs like left recursion and so on. To do so we can define queries which generate some errors or warnings (see section \ref{MoreQueries}). 

\begin{verbatim}
N {
    leftAssoc;
    rightAssoc;
};
{
    error on N : "A symbol cannot be left- and 
               right-associative at the same time";
}
\end{verbatim}

In this example we prohibit using attributes ``leftAssoc'' and ``rightAssoc'' on the same symbol at the same time. These checks may be put into aspects along with metadata attachment rules which makes diagnostic modular and reusable. 
\section{Conclusion}

As we showed above Grammatic can solve many issues about grammar definition reuse and modularity. This tool is intended to be an open system so we are going to extend it by adding support for more and more external tools. Grammatic's generalized grammar definition format may serve as an interchange format for different tools which will again improve end user's experience.

\bibliographystyle{unsrt}
\bibliography{grammatic_main}

\begin{thebibliography}{10}

\bibitem{LanguageOriented}
Martin~P. Ward.
\newblock Language-oriented programming.
\newblock {\em Software --- Concepts and Tools}, 15(4):147--161, 1994.

\bibitem{KLV05}
Paul Klint, Ralf L\"{a}mmel, and Chris Verhoef.
\newblock Toward an engineering discipline for grammarware.
\newblock {\em ACM Trans. Softw. Eng. Methodol.}, 14(3):331--380, 2005.

\bibitem{yacc}
Steven~C. Johnson.
\newblock Yacc: Yet another compiler compiler.
\newblock In {\em {UNIX} Programmer's Manual}, volume~2, pages 353--387. Holt,
  Rinehart, and Winston, New York, NY, USA, 1979.

\bibitem{Parr2007}
Terence Parr.
\newblock {\em The Definitive ANTLR Reference: Building Domain-Specific
  Languages}.
\newblock The Pragmatic Bookshelf, Raleigh, 2007.

\bibitem{Mossenbock05thecompiler}
Hanspeter Mössenböck and Johannes Kepler.
\newblock The compiler generator {Coco/R} -- user manual, 2005.

\bibitem{oAWxText}
{Open Architecture Ware}.
\newblock {xText}.
\newblock \\http://www.openarchitectureware.org/, 2007.

\bibitem{JouaultBezKur2006}
Fr{\`e}d{\`e}ric Jouault, Jean B{\`e}zivin, and Ivan Kurtev.
\newblock {TCS: a DSL for the Specification of Textual Concrete Syntaxes in
  Model Engineering}.
\newblock In {\em GPCE 2006}, ACM, pages 249--254, 2006.

\bibitem{Gagnon98anobject-oriented}
Etienne Gagnon and Laurie Hendren.
\newblock An object-oriented compiler framework.
\newblock In {\em In Proceedings of TOOLS}, pages 140--154, 1998.

\bibitem{BDHJJKKMOSVVV01}
J.~Heering, P.~R.~H. Hendriks, P.~Klint, and J.~Rekers.
\newblock The syntax definition formalism {SDF} -- reference manual.
\newblock {\em SIGPLAN Not.}, 24(11):43--75, 1989.

\bibitem{Mernik00compiler}
Marjan Mernik, Mitja Leni, and Enis A.
\newblock Compiler/interpreter generator system {LISA}.
\newblock In {\em In IEEE Proceedings of 33rd Hawaii International Conference
  on System Sciences}, pages 590--594, 2000.

\bibitem{Parr08}
Terence Parr.
\newblock The reuse of grammars with embedded semantic actions.
\newblock {\em International Conference on Program Comprehension}, 0:5--10,
  2008.

\bibitem{Kiczales97aspect-orientedprogramming}
Gregor Kiczales, John Lamping, Anurag Mendhekar, Chris Maeda, Cristina~Videira
  Lopes, Jean marc Loingtier, John Irwin, Gregor Kiczales, John Lamping, Anurag
  Mendhekar, Chris Maeda, Cristina Lopes, Jean marc Loingtier, and John Irwin.
\newblock Aspect-oriented programming.
\newblock In {\em In ECOOP}. SpringerVerlag, 1997.

\bibitem{680006}
Gregor Kiczales, Erik Hilsdale, Jim Hugunin, Mik Kersten, Jeffrey Palm, and
  William~G. Griswold.
\newblock An overview of {AspectJ}.
\newblock In {\em ECOOP '01: Proceedings of the 15th European Conference on
  Object-Oriented Programming}, pages 327--353, London, UK, 2001.
  Springer-Verlag.

\end{thebibliography}
\end{document}